\documentclass[a4paper]{jpconf}
\usepackage{amssymb,amsfonts,amsmath}
\usepackage{graphicx}
\usepackage{subfigure}
\usepackage{hyperref}
\usepackage[numbers,square,sort&compress]{natbib}

\begin{document}
\title{Early Advanced LIGO binary neutron-star sky localization and parameter estimation}

%\author{C~P~L~Berry$^1$, B~Farr$^2$, W~M~Farr$^1$, C-J~Haster$^1$, I~Mandel$^1$, H~Middleton$^1$, L~P~Singer$^3$, A~L~Urban$^4$, A~Vecchio$^1$, S~Vitale$^5$, K~Cannon$^6$, P~B~Graff$^{3,7}$, C~Hanna$^{8,9}$, S~Mohapatra$^{5,10}$, C~Pankow$^4$, L~R~Price$^{11}$, T~Sidery$^1$ and J~Veitch$^1$}
\author{C~P~L~Berry$^1$, B~Farr$^2$, W~M~Farr$^1$, C-J~Haster$^1$, I~Mandel$^1$, H~Middleton$^1$, L~P~Singer$^3$, A~L~Urban$^4$, A~Vecchio$^1$, S~Vitale$^5$, K~Cannon$^6$, P~B~Graff$^{7,8}$, C~Hanna$^{9,10}$, S~Mohapatra$^{5,11}$, C~Pankow$^4$, L~R~Price$^{12}$, T~Sidery$^1$ and J~Veitch$^1$}

\address{$^1$ School of Physics \& Astronomy, University of Birmingham, Birmingham, B15 2TT, UK}
\address{$^2$ Enrico Fermi Institute, University of Chicago, Chicago, IL 60637, USA}
%\address{$^3$ NASA Goddard Space Flight Center, Greenbelt, MD 20771, USA}
\address{$^3$ Astrophysics Science Division, NASA Goddard Space Flight Center, Greenbelt, MD 20771, USA}
\address{$^4$ Leonard E.\ Parker Center for Gravitation, Cosmology, and Astrophysics, University of Wisconsin--Milwaukee, Milwaukee, WI 53201, USA}
\address{$^5$ Massachusetts Institute of Technology, 185 Albany St, Cambridge, MA 02139, USA}
%\address{$^6$ CITA, 60 St.\ George Street, University of Toronto, Toronto, Ontario, M5S 3H8, Canada}
\address{$^6$ Canadian Institute for Theoretical Astrophysics, 60 St.\ George Street, University of Toronto, Toronto, Ontario, M5S 3H8, Canada}
\address{$^7$ Department of Physics, University of Maryland--College Park, College Park, MD 20742, USA}
%\address{$^8$ Perimeter Institute for Theoretical Physics, Ontario, N2L 2Y5, Canada}
%\address{$^{9}$ Pennsylvania State University, University Park, PA 16802, USA}
%\address{$^{10}$ Syracuse University, Syracuse, NY 13244, USA}
%\address{$^{11}$ LIGO Laboratory, California Institute of Technology, Pasadena, CA 91125, USA}
\address{$^8$ Gravitational Astrophysics Lab, NASA Goddard Space Flight Center, Greenbelt, MD 20771, USA}
\address{$^9$ Perimeter Institute for Theoretical Physics, Ontario, N2L 2Y5, Canada}
\address{$^{10}$ Pennsylvania State University, University Park, PA 16802, USA}
\address{$^{11}$ Syracuse University, Syracuse, NY 13244, USA}
\address{$^{12}$ LIGO Laboratory, California Institute of Technology, Pasadena, CA 91125, USA}

\ead{cplb@star.sr.bham.ac.uk}

\begin{abstract}
2015 will see the first observations of Advanced LIGO and the start of the gravitational-wave (GW) advanced-detector era. One of the most promising sources for ground-based GW detectors are binary neutron-star (BNS) coalescences. In order to use any detections for astrophysics, we must understand the capabilities of our parameter-estimation analysis. By simulating the GWs from an astrophysically motivated population of BNSs, we examine the accuracy of parameter inferences in the early advanced-detector era. We find that sky location, which is important for electromagnetic follow-up, can be determined rapidly ($\sim 5~\mathrm{s}$), but that sky areas may be hundreds of square degrees. The degeneracy between component mass and spin means there is significant uncertainty for measurements of the individual masses and spins; however, the chirp mass is well measured (typically better than $0.1\%$).
\end{abstract}

\section{Introduction}

The advanced generation of ground-based gravitational-wave (GW) detectors, Advanced LIGO (aLIGO) \cite{TheLIGOScientific:2014jea} and Advanced Virgo (AdV) \cite{TheVirgo:2014hva}, begin operation soon: the first observing run (O1) of aLIGO is September 2015--January 2016 \cite{Aasi:2013wya}. Binary neutron stars (BNSs) are a promising source \cite{Abadie:2010cf}.\footnote{Since submission, the first detection (of a binary black hole rather than a BNS), has been announced \cite{Abbott:2016blz}.}

Analysis of a signal goes through several stages: detection, low-latency parameter estimation (PE), mid-latency PE and high-latency PE \cite{WhitePaper2014}. Each refines our understanding. To discover what we can learn about BNSs, a simulated astrophysically motivated population of BNS signals (component masses $m_{1,2} \in [1.2, 1.6]M_\odot$, isotropic spins with magnitudes $a_{1,2} \in [0, 0.05]$, and uniformly distributed in volume \cite{Singer:2014qca}) has been studied in an end-to-end analysis, with results reported in several publications. Singer \textit{et al}.~\cite{Singer:2014qca} studied the (low- and mid-latency) prospects for sky localization.\footnote{Singer \textit{et al}.~\cite{Singer:2014qca} also considered the second observing run (O2), with AdV joining the network.}
Berry \textit{et al}.~\cite{Berry:2014jja} repeated the analysis using more realistic noise (detector noise from the sixth science run of initial LIGO \cite{Aasi:2014usa} recoloured to match the expected sensitivity of early aLIGO \cite{Barsotti:2012}), in contrast to ideal Gaussian noise. In addition to considering sky localization, Berry \textit{et al}.~\cite{Berry:2014jja} also investigated measurements of source distance and mass. The latter is influenced by spin, Farr \textit{et al}.~\cite{Farr:2015lna} completed the high-latency analysis including the effects of spin, considering all aspects of PE. We report results from these studies for O1 PE; further technical details are in the papers themselves. %In Sec.~\ref{sec:sky} we look at sky localization, and in Sec.~\ref{sec:mass} we look at mass and spin.

\section{Sky localization}\label{sec:sky}

%A BNS merger could be accompanied by an electromagnetic counterpart \cite{Metzger:2011bv}; performing EM follow-up of a GW trigger requires the knowledge of the source sky location. With a two-detector network, timing triangulation produces a band on the sky, a constant angle from the line joining the two detectors \cite{Fairhurst:2009tc,Fairhurst:2010is}. The degeneracy along this circle is broken by amplitude and phase information; however, localization still follows extended arcs \cite{Singer:2014qca}.

Sky localization can be computed at low-latency by \textsc{bayestar} \cite{Singer:2015ema} or at mid- to high-latency by \textsc{LALInference} \cite{Veitch:2014wba}.\footnote{Part of the LIGO Algorithm Library (LAL) available from \href{http://www.lsc-group.phys.uwm.edu/lal}{www.lsc-group.phys.uwm.edu/lal}.} Both are fully Bayesian PE codes; \textsc{bayestar} uses the output of the detection pipeline, while \textsc{LALInference} matches GW templates to the measured detector strain \cite{Cutler:1994ys}. Computing templates is computationally expensive; mid-latency PE is done with (non-spinning) TaylorF2 and high-latency PE is done with (fully spin-precessing) SpinTaylorT4. Both are inspiral-only post-Newtonian waveforms \cite{Buonanno:2009zt}. \textsc{bayestar} takes a median time of $4.5~\mathrm{s}$ to calculate the location \cite{Singer:2015ema}; the median times for the non-spinning and spinning \textsc{LALInference} analyses to collect $2000$ posterior samples are $\sim5.7 \times 10^4~\mathrm{s}$ \cite{Berry:2014jja} and $\sim 9.2 \times 10^5~\mathrm{s}$ \cite{Farr:2015lna} respectively.

\begin{figure}
  \centering
   \subfigure[$50\%$ credible region]{\includegraphics[width=0.45\columnwidth]{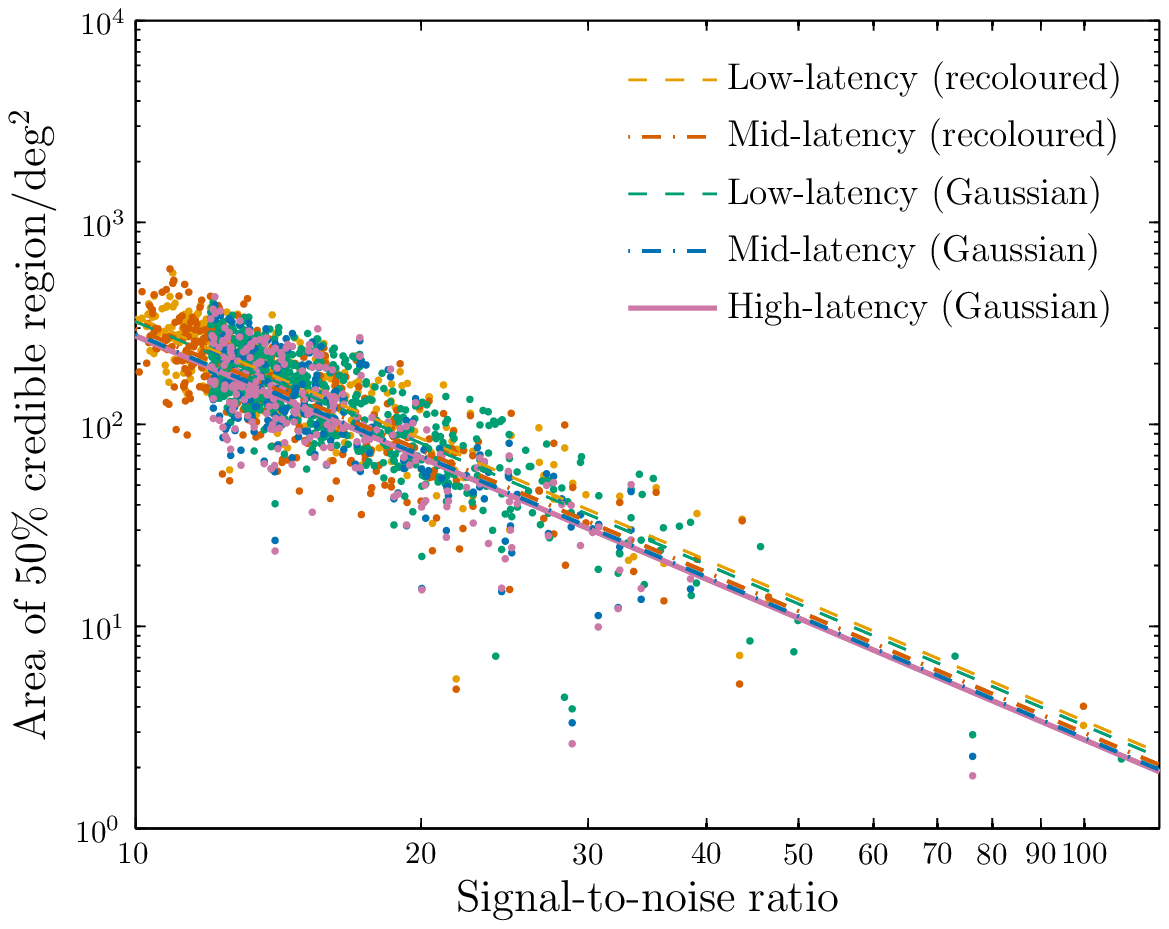}} \quad
   \subfigure[$90\%$ credible region]{\includegraphics[width=0.45\columnwidth]{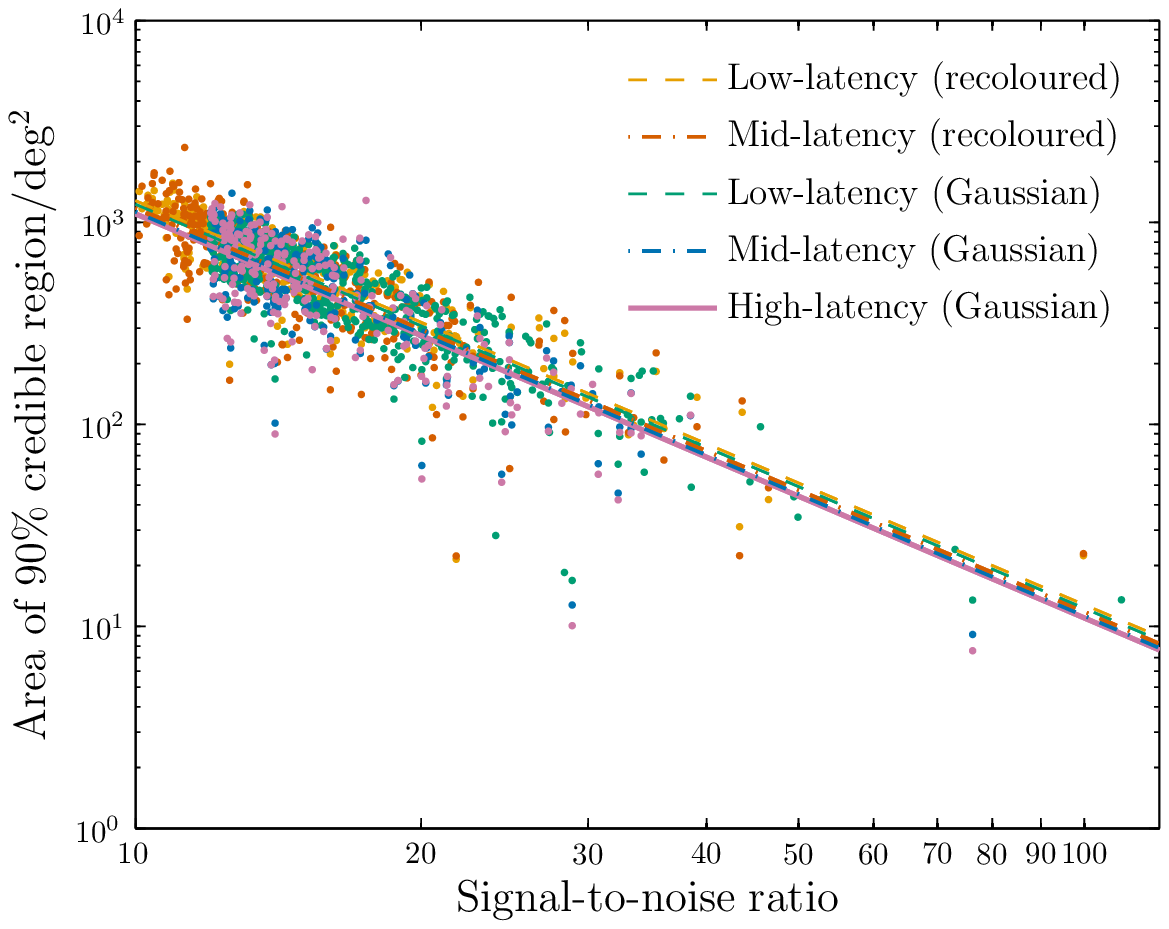}}
    \caption{Sky localization versus SNR for the low-latency \textsc{bayestar}, the mid-latency (non-spinning) \textsc{LALInference} and the high-latency (spinning) \textsc{LALInference} analyses \cite{Singer:2014qca,Berry:2014jja,Farr:2015lna}. Individual results are indicated by points and lines indicate best fits assuming $\mathrm{CR}_{p} \propto \varrho^{-2}$; these are $\mathrm{CR}_{0.5} \approx  (2.84\times 10^4) \varrho^{-2}~\mathrm{deg^{2}}$ and $\mathrm{CR}_{0.9} \approx (1.14\times 10^5) \varrho^{-2}~\mathrm{deg^{2}}$ across the range considered.} 
    \label{fig:snr-area}
\end{figure}

Despite their differences, \textsc{bayestar} and \textsc{LALInference} produce consistent results for a two-detector network.\footnote{This is not the case in a three-detector network if there is not a trigger from all the detectors \cite{Singer:2014qca,Singer:2015ema}.} The inclusion of spin in PE does not change sky localization for this slowly spinning population (the same may not be true for rapidly spinning black holes). At a constant signal-to-noise ratio (SNR) $\varrho$, there is also a negligible difference between results from Gaussian and recoloured noise. The scaling of the $50\%$ credible region $\mathrm{CR}_{0.5}$ and $90\%$ credible region $\mathrm{CR}_{0.9}$ with SNR is shown in Fig.~\ref{fig:snr-area}. Assuming a detection threshold of a false alarm rate of $10^{-2}~\mathrm{yr^{-1}}$ ($\varrho \gtrsim 10$--$12$), the median $\mathrm{CR}_{0.5}$ ($\mathrm{CR}_{0.9}$) is $170~\mathrm{deg^2}$ ($690~\mathrm{deg^2}$) using \textsc{bayestar} and $150~\mathrm{deg^2}$ ($630~\mathrm{deg^2}$) using \textsc{LALInference}; switching to a threshold of $\varrho \geq 12$ \cite{Aasi:2013wya}, these become $140~\mathrm{deg^2}$ ($520~\mathrm{deg^2}$) and $120~\mathrm{deg^2}$ ($480~\mathrm{deg^2}$) respectively \cite{Berry:2014jja}.

\section{Mass and spin}\label{sec:mass}

The first estimates for the component masses $m_{1,2}$ come from the detection pipeline, here \textsc{GSTLAL} \cite{Cannon:2011vi}. Full posteriors are constructed by \textsc{LALInference}. The degeneracy between mass and spin complicates measurements. Excluding spins (as in the mid-latency analysis) means we can miss the true parameter values. Allowing spins to vary over the full (black hole) range of $a_{1,2} \in [0, 1]$ (as in the high-latency analysis) and including precession ensures we cover the true value, but potentially means that we consider spin values not found in nature: here, the spins are $a_{1,2} < 0.05$, but we will not know the true distribution in practice.

\begin{figure}
  \centering
   \subfigure[Offset from true value\label{fig:offset}]{\includegraphics[width=0.45\columnwidth]{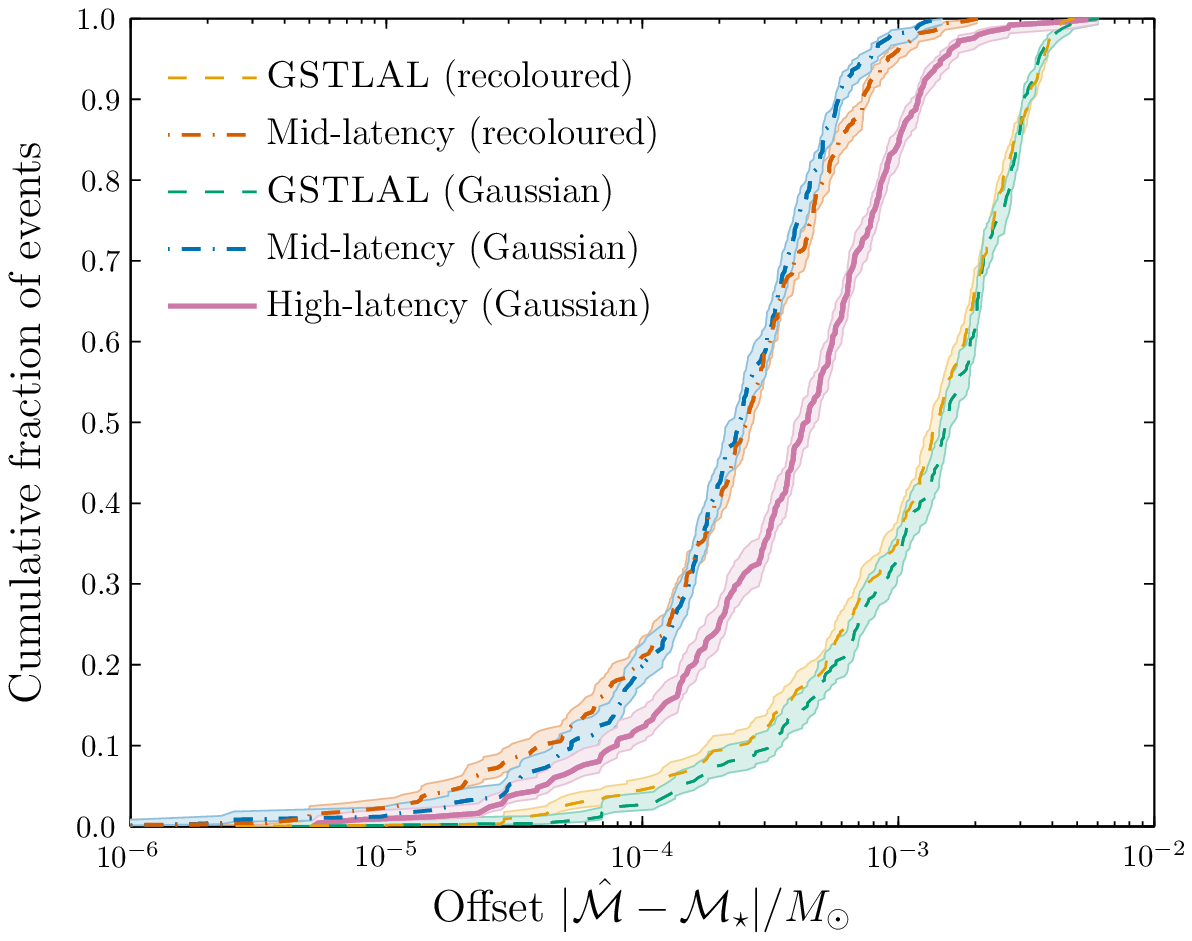}} \quad
   \subfigure[Posterior width\label{fig:sigma}]{\includegraphics[width=0.45\columnwidth]{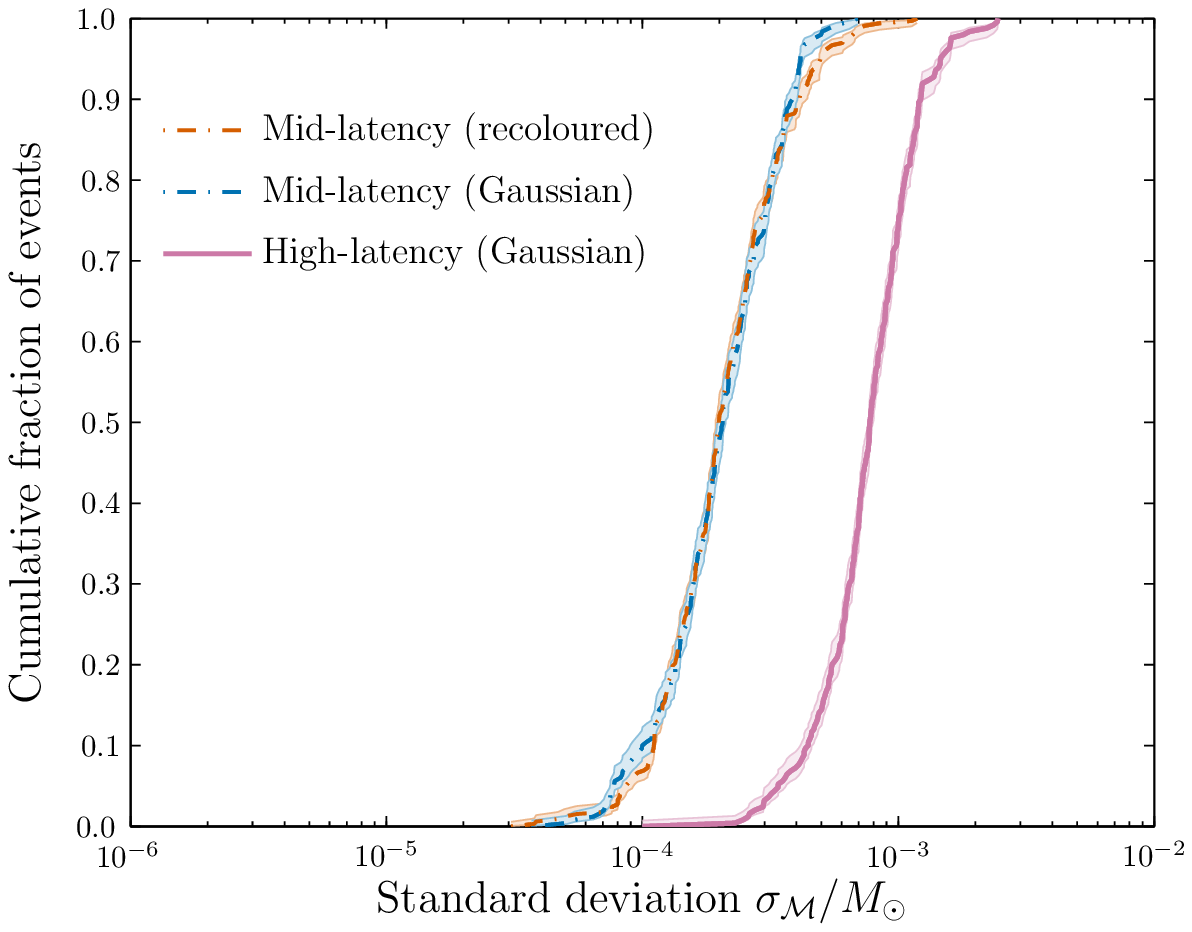}}
    \caption{Cumulative fractions of events with (a) offsets in chirp-mass estimates and (b) posterior standard deviations smaller than the abscissa value \cite{Berry:2014jja,Farr:2015lna}. The offset is the difference between the true value $\mathcal{M}_\ast$ and maximum likelihood value from \textsc{GSTLAL} or the posterior mean from (mid- or high-latency) \textsc{LALInference}. The shaded areas are the $68\%$ confidence intervals on the cumulative distributions.} 
    \label{fig:masses}
\end{figure}

The chirp mass $\mathcal{M} = (m_1 m_2)^{3/5}/(m_1+m_2)^{1/5}$ is the best measured mass parameter. Fig.~\ref{fig:offset} shows the offset between chirp-mass estimates (maximum likelihood values for \textsc{GSTLAL} and posterior means for \textsc{LALInference}) and the true values. All methods produce accurate results (offsets $<0.5\%$) and there is no noticeable difference between recoloured and Gaussian noise. The mid-latency offsets are smaller than the high-latency ones, because our BNSs are slowly spinning (which need not be the case in reality). However, the mid-latency offsets are more statistically significant. The mean values of $(\hat{\mathcal{M}} - \mathcal{M})^2/\sigma_\mathcal{M}^2$, where $\sigma_\mathcal{M}$ is the posterior standard deviation, are $5.5$, $5.1$ and $0.7$ for the recoloured non-spinning, Gaussian non-spinning and Gaussian spinning analyses respectively. Ignoring spin yields posteriors that are too narrow \cite{Berry:2014jja}, the distribution of $\sigma_\mathcal{M}$ is shown in Fig.~\ref{fig:sigma} \cite{Farr:2015lna}; the median values of $\sigma_\mathcal{M}$ are $2.0\times10^{-4}M_\odot$, $2.1\times10^{-4}M_\odot$ and $7.7\times10^{-4}M_\odot$ for the recoloured non-spinning, Gaussian non-spinning and Gaussian spinning analyses respectively.

Measurements of other mass parameters, such as the mass ratio $q = m_2/m_1$ ($0 < q \leq 1$) or $m_{1,2}$, are less precise, and the degeneracy with spin is more pronounced \cite{Cutler:1994ys,Farr:2015lna}: the median $50\%$ ($90\%$) credible interval for $q$ is $0.29$ ($0.59$). For our population of low-spin BNSs, the spins are not well measured and have large uncertainties. None of the events have a $50\%$ upper credible bound less than $0.1$; the median $50\%$ ($90\%$) upper credible bound is $0.30$ ($0.70$) for $a_1$ (the dominant spin) and $0.42$ ($0.86$) for $a_2$. Low spin values are preferred, but spin magnitudes can only be weakly constrained.

\section{Summary}

O1 marks the beginning of the advanced-detector era. As time progresses, sensitivities improve and further detectors (AdV, LIGO-India \cite{Indigo} and KAGRA \cite{Aso:2013eba}) come online, the prospects for detection and PE will become better \cite{Schutz:2011tw,Veitch:2012df,Singer:2014qca}. %\footnote{The exception may be distance, as improved sensitivity means we can observe sources from a greater range.}
For BNSs, chirp mass is always well measured, but sky localization and spins are more uncertain.

\ack
This work was supported in part by STFC. This is LIGO document reference LIGO-P1500155. A catalogue of results is available at \href{http://www.ligo.org/scientists/first2years/}{www.ligo.org/scientists/first2years/}.

%\bibliography{amaldi.bib}
\providecommand{\newblock}{}

\end{document}